\documentclass[aps,floatfix,prl,twocolumn,letterpaper,10pt,twoside,tightenlines,nofootinbib,showpacs]{revtex4}
\usepackage{graphicx}
\usepackage{color}

\begin{document}

\title{Heavy-Quark Probes of the Quark-Gluon Plasma at RHIC} 

\author{Hendrik van Hees$^{1}$, Vincenzo Greco$^{2}$ and Ralf Rapp$^{1}$} 

\affiliation{$^1$Cyclotron Institute and Physics Department, Texas A{\&}M
  University, College Station, Texas 77843-3366, USA \\
  $^2$Laboratori Nazionali del Sud INFN, via S. Sofia 62, I-95123 Catania, Italy}

\date{\today}

\begin{abstract}
  Thermalization and collective flow of charm ($c$) and bottom ($b$)
  quarks in ultra-relativistic heavy-ion collisions are evaluated based
  on elastic parton rescattering in an expanding quark-gluon plasma
  (QGP). We show that resonant interactions in a strongly interacting
  QGP (sQGP), as well as parton coalescence, can play an
  essential role in the interpretation of recent data from the
  Relativistic Heavy-Ion Collider (RHIC), and thus illuminate the nature
  of the sQGP and its hadronization. Our main assumption, motivated by
  recent findings in lattice Quantum Chromodynamics, is
  the existence of $D$- and $B$-meson states in the sQGP, providing
  resonant cross sections for heavy quarks. 
  Pertinent drag and diffusion coefficients are implemented into a
  relativistic Langevin simulation to compute transverse-momentum
  spectra and azimuthal asymmetries ($v_2$) of $b$- and $c$-quarks in
  Au-Au collisions at RHIC. After hadronization into $D$- and $B$-mesons
  using quark coalescence and fragmentation,
 associated electron-decay spectra and $v_2$ are compared to recent 
 RHIC data. Our results suggest a reevaluation of radiative and elastic 
 quark energy-loss mechanisms in the sQGP.
\end{abstract}

\maketitle

\textit{Introduction.}  Recent experimental findings at the Relativistic
Heavy-Ion Collider (RHIC) have given intriguing evidence for the
production of matter at unprecedented (energy-) densities with
surprisingly large collectivity and opacity, as reflected by
(approximately) hydrodynamic behavior at low transverse momentum ($p_T$)
and a substantial suppression of particles with high $p_T$. This has led
to the notion of a ``strongly interacting Quark-Gluon Plasma'' (sQGP),
whose microscopic properties, however, remain under intense debate thus
far.

Heavy quarks (HQs) are particularly valuable probes of the medium
created in heavy-ion reactions, as one expects their production to be
restricted to the primordial stages. Recent calculations of radiative
gluon energy-loss of charm ($c$) quarks traversing a QGP in central
Au-Au collisions at RHIC have found nuclear suppression factors
$R_{AA}$$\simeq$0.3-0.4~\cite{djo04,arm05}, comparable to the observed
suppression of light hadrons at high $p_T$, and in line with preliminary
(non-photonic) single-electron ($e^\pm$) decay
spectra~\cite{phenix-e1,jac05,bil05}. The latter also exhibit a
surprisingly large azimuthal asymmetry
($v_2$)~\cite{v2-phenix,v2pre-star,aki05} in semicentral Au-Au which cannot be
reconciled with radiative energy loss, especially if $c$-quarks are
hadronized into $D$-mesons via fragmentation. While the underlying
transport coefficients~\cite{arm05} exceed their predicted values from
perturbative Quantum Chromodynamics (pQCD) by at least a factor of
$\sim$5~\cite{bai02}, energy loss due to {\em elastic} scattering
parametrically dominates toward low $p_T$ (by a factor
1/$\sqrt{\alpha_s}$~\cite{MT04}). But elastic pQCD cross
sections~\cite{Svet88,MT03} also have to be upscaled substantially to
obtain $c$-quark $v_2$ and $R_{AA}$ reminiscent to preliminary $e^\pm$
data, as shown in a recent Langevin simulation for RHIC~\cite{MT04}.  In
addition, contributions of bottom ($b$) quarks~\cite{Cac05,djo05} will
reduce the effects in the electron-$R_{AA}$ and -$v_2$ at high $p_T$.
Quark coalescence approaches suggest that an $e^\pm$-$v_2$ in excess of
10\% can only be obtained if~\cite{GKR04} (i) light quarks impart their
$v_2$ on $D$-mesons (see also Refs.~\cite{Mol04,Zhang05}), (ii) the
$c$-quark $v_2$ is comparable to that of light quarks.

We are thus confronted with marked discrepancies between pQCD
energy-loss calculations and semileptonic heavy-quark (HQ) observables
at RHIC. The resolution of this issue is central to the understanding of
HQ interactions in the QGP in particular, and to the interpretation of
energy loss in general. HQ rescattering also has direct impact on other
key observables such as heavy quarkonium production (facilitating
regeneration) and dilepton spectra (where $c\bar c$ decays compete with
thermal QGP radiation).

In this letter we investigate the HQ energy-loss problem by introducing
resonant HQ interactions into a Langevin simulation of an expanding QGP.
Our calculations implement a combined
coalescence+fragmentation approach for hadronization, as well as bottom
contributions, to allow for a quantitative evaluation of pertinent
$e^\pm$-spectra ($v_2$ and $R_{AA}$) which is mandatory for a proper
interpretation of recent RHIC data.
Our main assumption of resonant $D$- and $B$-like states in the sQGP has
been shown~\cite{HR05a} to reduce HQ thermalization times by a
factor of $\sim$3 compared to pQCD scattering. Theoretical evidence for
resonances in the sQGP derives from computations of heavy and light
meson correlators within lattice QCD (lQCD)~\cite{AH-prl,KL03}, as well
as applications of lQCD-based heavy-quark potentials within effective
models~\cite{SZ04,Wong04,MR05,Alberico05}. Except for the mass and width
of these states, no further free parameters enter our description, with
degeneracies based on chiral and HQ symmetry.

\textit{Heavy-Quark Interactions in the QGP.}  Following
Ref.~\cite{HR05a} our description of HQ interactions 
in the QGP focuses on elastic scattering, mediated by resonance
excitations on light antiquarks ($\bar q$) as well as (nonresonant)
leading order pQCD processes dominated by $t$-channel gluon exchange.
The latter correspond to Born diagrams~\cite{com79} regularized by a
gluon-screening mass $m_g$=$g T$ with a strong coupling constant,
$\alpha_s$=$g^2/(4 \pi)$=0.4.  The key assumption~\cite{HR05a} is that a
QGP at moderate temperatures $T$$\le$2$T_c$ sustains strong correlations
in the lowest-lying color-neutral $D$- and $B$-meson channels. Support
for the relevance of such interactions stems from quenched lQCD
computations of euclidean mesonic correlation functions, which, after
transformation into the timelike regime, exhibit resonance structures
for both (heavy) $Q$-$\bar Q$ and (light) $q$-$\bar q$ 
states~\cite{AH-prl,KL03}.  In
addition, applications of lQCD-based $Q$-$\bar Q$ potentials have
revealed both bound~\cite{SZ04,Wong04} and resonance states~\cite{MR05}
with dissolution temperatures of $\sim2T_c$, quite compatible with the
disappearance of the peak structures in the lQCD spectral functions.
Here, we do not attempt a microscopic description of these correlations
but cast them into an effective lagrangian with $\bar q$-$Q$-$\Phi$
vertices ($\Phi$=$D$, $B$), at the price of 2 free parameters: the
masses of the meson-fields, fixed at $m_{D(B)}$=2(5)~GeV, {\it i.e.},
0.5~GeV {\it above} the $Q$-$\bar q$ threshold (with quark masses
$m_{c(b)}$=1.5(4.5)~GeV, $m_{u,d}$=0),
and their width, $\Gamma$, obtained from the one-loop $\Phi$
self-energy (which, in turn, dresses the $\Phi$-propagator) with the
pertinent coupling constant varied to cover a range suggested by
effective quark models~\cite{GK92,Blasch03}, $\Gamma$=0.4-0.75~GeV. The
multiplicity of $\Phi$ states follows from chiral and HQ symmetries
alone, implying degenerate $J^P$=$0^\pm$ and 1$^\pm$ states.  We
emphasize again that, besides the mass and width of the $\Phi$ states,
no other free parameters (or scale factors) are introduced.

The matrix elements for resonant and pQCD scattering are employed to
calculate drag and diffusion coefficients of HQs in a Fokker-Planck
approach~\cite{Svet88}. Resonances reduce the thermalization times for
both $c$- and $b$-quarks by a factor of $\sim$$3$ compared to pQCD
scattering alone~\cite{HR05a}.

\textit{Langevin Simulation.}  To evaluate thermalization and collective
flow of HQs in Au-Au collisions we perform relativistic Langevin
simulations~\cite{MT04} embedded into an expanding QGP fireball.  In the
local rest frame of the bulk matter, the change in position ($\vec x$)
and momentum ($\vec p$) of $c$- and $b$-quarks during a time step
$\delta t$ is defined by
\begin{equation}
\label{langevin}
\delta \vec{x}=\frac{\vec{p}}{E}~\delta t \ , \quad
\delta \vec{p}=-A(t,\vec{p}+\delta \vec{p})~\vec{p}~\delta t + \delta
\vec{W}(t,\vec{p}+\delta \vec{p}) \ 
\end{equation}
($E$: HQ energy), where $\delta \vec{W}$ represents a random force which
is distributed according to Gaussian noise~\cite{hae05},
\begin{equation}
\label{fluct-force}
P(\delta \vec{W}) \propto \exp \left [-\frac{\hat{B}_{jk} \delta W^j \delta
    W^{k}}{4 \delta t} \right] \ .
\end{equation}
The drag coefficient (inverse relaxation time), $A$, and the inverse of
the diffusion-coefficient matrix,
\begin{equation}
B_{jk}=B_{0} (\delta^{jk} - \hat{p}^j \hat{p}^k)
+ B_1 \hat{p}^j \hat{p}^k \ ,  
\end{equation}
are given by the microscopic model of Ref.~\cite{HR05a}, including $p$-
and $T$-dependencies (the latter converts into a time dependence using
the fireball model described below). The longitudinal diffusion
coefficient is set in accordance with Einstein's 
dissipation-fluctuation relation to~\cite{MT04}
$B_1$=$TEA$,  
to ensure the proper thermal equilibrium limit. The latter also requires
care in the realization of the stochastic process in
Eq.~(\ref{langevin}); we here use the so-called H\"anggi-Klimontovich
realization~\cite{hae05} which approaches a relativistic Maxwell
distribution if the Einstein relation
is satisfied (in the Ito realization, {\it e.g.}, an extra
term has to be introduced in $A$~\cite{arnold}). Finally, the HQ momenta
are Lorentz-boosted to the laboratory frame with the velocity of the
bulk matter at the actual position of the quark, as determined by the 
fireball flow profile (see below).

The time evolution of Au-Au collisions is modeled by an
isentropically expanding, isotropic QGP fireball with a total entropy
fixed to reproduce measured particle multiplicities at hadro-chemical
freezeout which we assume to coincide with the phase transition at
$T_c$=180~MeV~\cite{Ra01}. The temperature at each instant of time is
extracted from an ideal QGP equation of state with an effective flavor
degeneracy of $N_f$=2.5. Radial and elliptic flow of the bulk matter
are parameterized to closely resemble the time-dependence found in
hydrodynamical calculations~\cite{kol00}, assuming a 
flow profile rising linearly with the radius. We focus on semicentral
collisions at impact parameter $b$=7~fm, with an initial spatial
eccentricity of 0.6 and a formation time of $\tau_0$=0.33~fm/c,
translating into an initial temperature of $T_0$=340~MeV. The evolution
is terminated at the end of the QGP-hadron gas mixed phase (constructed
via standard entropy balance~\cite{Ra01}) after about 5~fm/c, at which
point the surface flow velocity and momentum anisotropy have reached
$v_\perp$=0.5c and $v_2$=5.5\% (variations in $\tau_0$
by a factor of two affect the $c$-quark $v_2$ and $R_{AA}$ by 10-20\% 
(less for the $e^\pm$ spectra), while a reduction in the critical 
temperature to 170~MeV increases (decreases) $v_2^c$ ($R_{AA}$) 
somewhat less.  $D$-meson rescattering in the hadronic 
phase~\cite{fuchs04} is neglected).

To specify initial HQ $p_T$-distributions, $P_{\text{ini}}(p_{T})$, and
especially the relative magnitude of $c$- and $b$-quark spectra
(essential for the evaluation of $e^\pm$ spectra), we proceed as
follows: we first use modified PYTHIA $c$-quark spectra with
$\delta$-function fragmentation to fit $D$ and $D^*$ spectra in d-Au
collisions~\cite{star-D}.  These spectra are decayed to single-$e^\pm$
which saturate pertinent data from $p$-$p$ and d-Au up to
$p_T^e$$\simeq$3.5~GeV~\cite{phenix-pp,star-dAu}.  The missing yield at
higher $p_T$ is attributed to contributions from $B$-mesons, resulting
in a cross section ratio of $\sigma_{b\bar b}/\sigma_{c\bar
  c}$$\simeq$5$\cdot 10^{-3}$ (which is slightly smaller than for pQCD
predictions~\cite{Cac05} and implies a crossing of $c$- and $b$-decay
electrons at $p_T$$\simeq$5~GeV, as compared to 4~GeV in pQCD).

\begin{figure}[!tbh]
\centerline{\includegraphics[width=0.35\textwidth]{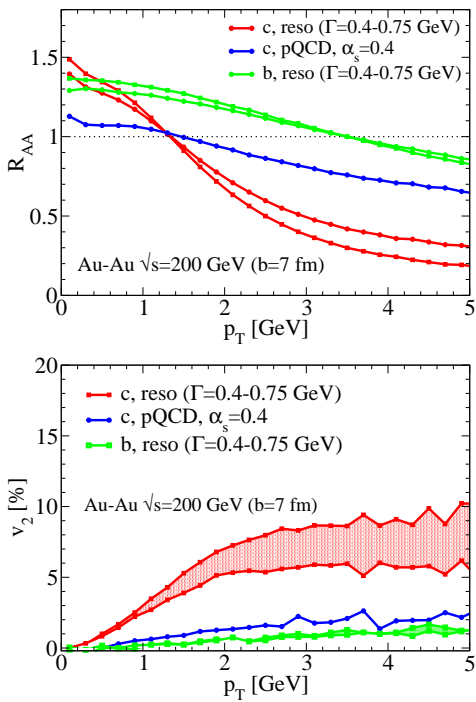}}
\caption{Nuclear modification factor (upper panel) and elliptic flow
  (lower panel) of charm and bottom quarks in $b$=7~fm
  Au-Au($\sqrt{s}$=200~GeV) collisions based on elastic rescattering in
  the QGP. Red (green) and blue lines are for $c$- ($b$-) quarks with
  and without resonance rescattering, respectively, where the bands
  encompass resonance widths of $\Gamma$=0.4-0.75~GeV.}
\label{fig1}
\end{figure}
Fig.~\ref{fig1} summarizes the output of the Langevin simula\-tions for
the HQ nuclear modification factor,
$R_{AA}$=$P_{\text{fin}}(p_{T})/P_{\text{ini}}(p_{T})$
($P_{\text{fin}}$: final $p_T$-distributions), and elliptic flow,
$v_2$=$\langle {(p_x^2-p_y^2)/(p_x^2+p_y^2)}_{p_T} \rangle$ (evaluated at fixed
$p_T$).  For $c$-quarks and pQCD scattering only, our results are in
fair agreement with those of Ref.~\cite{MT04} (recall that the Debye
mass in our calculations is given by $m_g$=$gT$ while in
Ref.~\cite{MT04} it was set to $m_g$=1.5$T$ independent of $\alpha_s$).
However, both $R_{AA}$ and $v_2$ exhibit substantial sensitivity to the
inclusion of resonance contributions, increasing the effects of pQCD
scattering by a factor of $\sim$3-5. Also note the development of the 
plateau in $v_2$($p_T$$>$3~GeV) characteristic for incomplete 
thermalization of HQs in the bulk matter.

\textit{Hadronization and Single-Electron Spectra.} 
\begin{figure}[!h]
\centering{\includegraphics[width=0.35\textwidth]{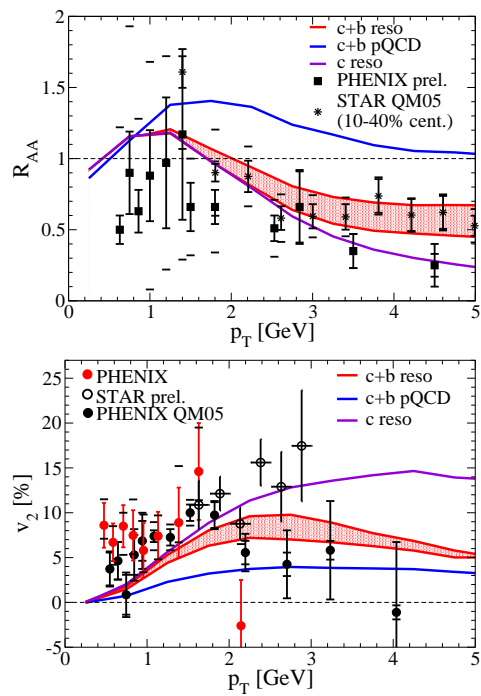}}
\caption{$R_{AA}$ (upper panel) and $v_2$ (lower panel) of semileptonic 
  $D$- and $B$-meson decay-electrons in $b$=7~fm
  Au-Au($\sqrt{s}$=200~GeV) using HQ coalescence supplemented by
  fragmentation (line identification as in Fig~\ref{fig1}). The
  data~\cite{jac05,v2-phenix,v2pre-star,aki05} are for minimum bias
  collisions, the $R_{AA}$ from STAR~\cite{bil05} for 10-40\% central collisions.}
  \label{fig2}
\end{figure}
Semi\-leptonic single-$e^\pm$ spectra are a
valuable tool to investigate heavy-meson spectra in ultrarelativistic
heavy-ion collisions, since their decay kinematics largely conserves the
spectral properties of the parent particles~\cite{GKR04,Dong:2004ve}.
To compare our results to  measured single-$e^\pm$ in Au-Au
collisions, the above HQ spectra have to be hadronized. To this end we
employ the coalescence approach of Ref.~\cite{GKR04} based on earlier
constructed light-quark spectra~\cite{Greco:2003mm}.  Quark coalescence
has recently enjoyed considerable success in describing, {\it e.g.}, the
``partonic scaling" of elliptic flow and the large $p$/$\pi$ ratio in
Au-Au at RHIC~\cite{Hwa:2002tu,Greco:2003mm,Fries:2003kq}, as well as
flavor asymmetries in $D$-meson production in elementary hadronic
collisions~\cite{Rapp:2003wn}. Whereas at low $p_T$ most of the HQs
coalesce into $D$- and $B$-mesons, this is no longer the case at higher
$p_T$ where the phase space density of light quarks rapidly decreases.
Therefore, to conserve HQ number in the $B$- and $D$-meson spectra, the
remaining $c$- and $b$-quarks are hadronized using $\delta$-function
fragmentation. Finally, single-$e^\pm$ $p_T$- and $v_2$-spectra are
computed via $B$- and $D$-meson 3-body decays, and compared to
experiment in Fig.~\ref{fig2}. We find that the effects of resonances
are essential in improving the agreement with data, both in terms of
lowering the $R_{AA}$ and increasing $v_2$. The $B$-meson contribution
reflects itself by limiting $R_{AA}$ and $v_2$ to values above 0.4 and
below 10\%, respectively, as well as the reduction of $v_2$ above
$p_T$$\simeq$3~GeV.

\begin{figure}[!tbh]
  \centering{\includegraphics[width=0.35\textwidth]{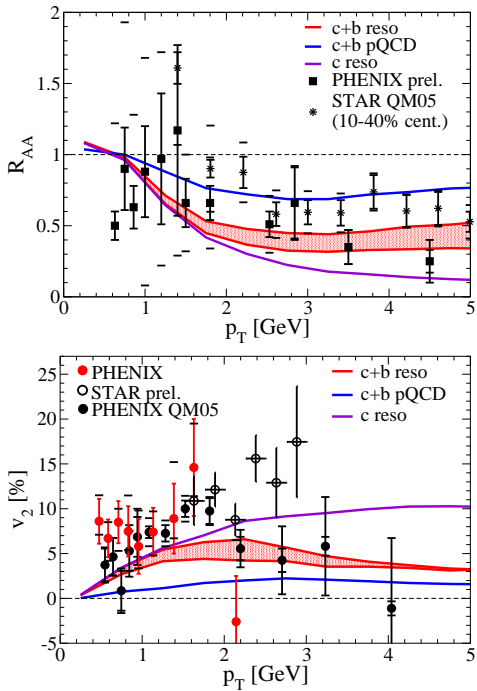}}
\caption{Same as Fig.~\ref{fig2} but using HQ fragmentation only.} 
\label{fig3}
\end{figure}

To better illustrate the effects of coalescence we plot in
Fig.~\ref{fig3} calculations where {\em all} HQs are fragmented into
$D$- and $B$-mesons.
While $R_{AA}$ is significantly reduced, most notably in the
$p_T$$\simeq$1-2~GeV region, $v_2$ also decreases reaching at most 6\%,
which is not favored by current data. It is, however, conceivable that
modifications in the fraction of coalescence to fragmentation
contributions, as well as improvements in our schematic
($\delta$-function) treatment of fragmentation, will be necessary once
more accurate experimental data become available.  Additional
corrections may also arise from a more precise determination of the
$b$/$c$ ratio and nuclear shadowing.

\textit{Conclusion.} We have investigated thermalization and collective
flow of $c$- and $b$-quarks within a relativistic Langevin approach
employing elastic scattering in an expanding QGP fireball in semicentral
Au-Au collisions at RHIC. Underlying drag and diffusion coefficients
were evaluated assuming resonant $D$- and $B$-meson correlations in the
sQGP, enhancing heavy-quark rescattering. Corresponding $p_T$-spectra
and elliptic flow of $c$-quarks exhibit a large sensitivity to the
resonance effects, lowering $R_{AA}$ down to 0.2 and raising $v_2$ up to
10\%, while the impact on $b$-quarks is small. Heavy-light quark
coalescence in subsequent hadronization significantly amplifies the
$v_2$ in single-electron decay spectra, but also increases their
$R_{AA}$, especially in the $p_T$$\simeq$2~GeV region. Bottom contributions
dominate above 3.5~GeV reducing both suppression and elliptic flow.  The
combined effects of coalescence and resonant heavy-quark interactions
are essential in generating a $v_2^e$ of up to 10\%, together with
$R_{AA}^e$$\simeq$0.5, supplying a viable explanation of current
electron data at RHIC without introducing extra scale factors.
Our analysis thus suggests that elastic rescattering of heavy quarks in
the sQGP is an important component for the understanding of heavy-flavor
and single-electron observables in heavy-ion reactions at collider
energies. While induced gluon-radiation is expected to be the prevalent
interaction with the medium at sufficiently high
$p_T$~\cite{djo04,arm05,djo05}, it may not be the dominant effect below
$p_T$$\simeq$6~GeV or so. A complete picture should clearly include
{\it both} elastic and inelastic rescattering mechanisms.

\textit{Acknowledgments:} One of us (HvH) thanks the Alexander von
Humboldt foundation for support within a Feodor Lynen fellowship. This
work was supported in part by a U.S. National Science Foundation CAREER
award under grant PHY-0449489.

\begin{flushleft}

\end{flushleft}

\end{document}